# Genome assembly in the telomere-to-telomere era

Heng Li[1,2,†], Richard Durbin[3,†]
[1] Department of Data Science, Dana-Farber Cancer Institute, Boston, MA, USA
[2] Department of Biomedical Informatics, Harvard Medical School, Boston, MA, USA
[3] Department of Genetics, Cambridge University, Cambridge, UK

[†]e-mail: hli@ds.dfci.harvard.edu and rd109@cam.ac.uk

**Abstract |** De novo assembly is the process of reconstructing the genome sequence of an organism from sequencing reads. Genome sequences are essential to biology, and assembly has been a central problem in bioinformatics for four decades. Until recently, genomes were typically assembled into fragments of a few megabases at best but technological advances in long-read sequencing now enable near complete chromosome-level assembly, also known as telomere-to-telomere assembly, for many organisms. Here we review recent progress on assembly algorithms and protocols. We focus on how to derive near telomere-to-telomere assemblies and discuss potential future developments.

## [H1] Introduction

Current sequencing technologies typically produce contiguous sequence reads of a hundred basepairs (bp) to a few tens of kilobases (kb) in length, very rarely more than one megabase (Mb). In contrast, most chromosomes in multi-cellular organisms are more than 10 Mb long, and they can be gigabases (Gb). To obtain the genome sequence of an organism, we need to obtain sequence reads that cover the genome multiple times and piece them together based on overlaps between them. The process of reconstructing the genome from sequence reads is called de novo assembly. In this review, we will often abbreviate this just to assembly for simplicity.

Historically there were two major strategies for whole-genome assembly: hierarchical sequencing and whole-genome shotgun sequencing (WGS). With the first strategy, we first tile the genome in cloned fragments of tens to hundreds of kilobases, then sequence and assemble the sequences of each clone, confirming the order of clone sequences with a physical or genetic map of the genome. *C. elegans*, the first assembled multi-cellular genome, was assembled this way[1]. The commonly used human reference genome GRCh38[2] was mainly derived from clone-based sequencing as well[3]. However, constructing and maintaining a comprehensive clone library is costly and labor intensive, and this approach is rarely if ever used today. With the second strategy, WGS, we randomly shear a genome into fragments, sequence these fragments and then reconstruct the genome[4,5]. Because we have to consider possible overlaps between reads from across the whole genome rather than between smaller sets of reads from relatively short localized clones, WGS-based assembly is more computationally challenging than clone-based assembly. Nonetheless, with improved assembly algorithms and data quality, WGS has become the dominant sequencing strategy for genome assembly.

The three critical properties of sequencing reads for assembly are length, accuracy and evenness of representation. Many of the issues that arise in designing assembly strategies involve tradeoffs between these.

The reference genomes of many model organisms were initially assembled about 20 years ago using Sanger reads of a few hundred to a thousand basepairs in length. Although high-throughput short reads are orders of magnitude cheaper to produce[6], they led to more fragmented assemblies owing to the short read lengths. The advent of single molecule sequencing allowed much longer reads that marked a turning point in sequence assembly. With noisy long reads available in 2010s, we could routinely assemble complete bacterial genomes[7] and start to automate human assemblies of reference quality[8]. Pure short-read assemblies are no longer competitive with assemblies that use long reads, in terms of continuity and contiguity.

However, at the end of 2019, we still could not assemble the entire genome of most multi-cellular organisms. Furthermore, almost all eukaryotes have a diploid or polyploid genome, consisting of two or more sets of haploid genomes in each living individual. In general, unless the organism is inbred as with laboratory models or otherwise experimentally manipulated, these haploid copies are similar but not identical. Assemblers developed up until 2020 either collapsed the sets of haploid genomes, losing 50% or more of genomic sequences, or produced fragmented assemblies of much lower quality than the reference genomes.

The arrival of accurate long reads[9] in 2019 has revolutionized the field of sequence assembly. With recent assemblers leveraging the high accuracy, we can routinely generate high-quality reference genome assemblies for new non-model research species[10–16]. An average human assembly we automatically produce today matches or exceeds the human reference genome GRCh38 which benefited from 20 years of curation efforts[17]. This enables whole genome de novo assembly of population samples[17,18] and leads to systematic projects to sequence entire groups of species, including ultimately all eukaryotes[19]. Now we have finished a human genome for the first time with each chromosome complete from telomere to telomere (T2T)[20]. We expect the same to be achieved for a rapidly increasing number of species in years to come.

In this article, we will review the current practices in the high-quality assembly of large eukaryotic genomes, ultimately towards finished telomere-to-telomere assembly. We will describe common data types, dissect recent assembly algorithms, explain methods for evaluating assemblies and discuss open challenges and future directions at the end.

## [H1] Properties of genomes that affect assembly

The main determinant of how easy it is to assemble a genome is not its size but its repeat structure (**Box 1**). A repetitive sequence can be resolved by reads longer than the repeat. However, there are longer repetitive regions. The pericentromeric region of human chromosome 1, for example, harbors 20 Mb of repeats[21], much longer than reads produced by current sequencing technologies. Nonetheless, we can still assemble this region with accurate long reads. Although it and other such regions are composed of similar repeat copies, they have accumulated mutations over time and rarely share an identical repeat sequence over 10 kb. Given long error-free reads, we can distinguish different repeat copies and successfully assemble them. Reads are never all entirely error-free, but when the read error rate is low enough and sequencing errors are sufficiently independent, we can correct most errors and achieve high-quality assembly.

The two homologous haplotypes in a diploid sample can also be viewed as repeats of each other. Correctly separating these two copies (or more than two in the case of polyploids) is known as "phasing". For a diploid or polyploid sample, a telomere-to-telomere assembly also implies all chromosomes are correctly phased. Phasing haplotypes and assembling repeats are related problems. An assembler capable of resolving similar repeats naturally has a high power to separate homologous haplotypes. Conversely, an assembler incapable of haplotype phasing is unable to resolve similar repeat copies. While traditional assembly algorithms collapse homologous haplotypes, current practices often preserve haplotype phasing over megabases, and can produce chromosome-scale haplotype-resolved assembly given multiple data types.

## [H1] Long-read and long-range sequencing technologies

Deriving a near telomere-to-telomere assembly often requires multiple sequencing technologies (**Table 1**). A long-read technology produces contiguous read sequences typically of ≥10 kb in length. Pacific Biosciences (PacBio) and Oxford Nanopore Technologies (ONT) are the two companies leading the development of long-read technologies. In 2019, PacBio introduced High-Fidelity (HiFi) reads that are 10–20 kb in length with an error rate below 0.5%. These have effectively replaced PacBio's older Continuous Long Reads (CLR) of >10% error rate. At present, HiFi reads are the core data type for high-quality assembly.

At the time of this review, ONT products available to the mass market have an accuracy broadly around 90–95%. For high-quality assembly, an important ONT data type is ultra-long reads that can be ≥100 kb in length. Despite their lower accuracy, these help to resolve remaining repetitive sequences that could not be assembled by HiFi reads. The most recent ONT v14 chemistry can generate simplex reads of 98–99% accuracy with the latest Dorado basecaller. ONT is actively developing duplex sequencing which sequences both strands of a DNA fragment. ONT duplex data approaches PacBio HiFi in accuracy and can be much longer. It will become a compelling data type once the technology matures.

Even ultra-long reads rarely span more than a few hundred kilobases. To reliably obtain chromosome-long scaffolds and phasing, we need long-range data. The most widely used long-range data type is Hi-C[22], which comprises short read pairs whose two ends can come from distant locations on the same chromosome. These provide phasing and contig ordering information over megabases. Pore-C[23] is similar to Hi-C but sequenced with ONT. Strand-seq[24] is another technology particularly good at chromosome grouping and contig orientation. However, it is more expensive and is not commercially available. Parental sequence data, or trio data, is powerful for whole-genome phasing[25,26] and can also be considered as a type of long-range data.

Linked-read technologies, including stLFR[27], TELL-seq[28] and haplotagging[29], produce clusters of short reads that come from fragments of the genome of ~100 kb. They require little DNA input and are cheap to produce. BioNano optical restriction digest maps provide long-range information as well. However, these technologies are not as powerful as ONT ultra-long reads or Hi-C in conjunction with HiFi contigs and are not often used in assembly.

## [H1] Recipes for near telomere-to-telomere assembly

While the first finished human chromosome (chromosome X) was assembled with ONT ultra-long reads[30,31], the whole human genome could only be completed with the advent of PacBio HiFi reads[20]. Current efforts towards telomere-to-telomere assembly of diploid samples focus on accurate long reads in combination with ONT ultra-long, trio and Hi-C data. It is generally observed that ≥15-fold read coverage per haplotype is necessary for deriving a contiguous assembly. Even thin ultra-long data of >100 kb at ~10-fold coverage noticeably improves the assembly, though higher coverage is still preferred. 30-fold Hi-C or trio coverage is usually sufficient.

Having explained above that assembly of most eukaryotic genomes from a diploid or polyploid individual involves phasing two or more sets of homologous chromosomes, below we will first consider the case where these sets are identical, or homozygous, so that we are effectively only assembling a single haploid genome. We then proceed to the next simplest case of heterozygous diploid samples. Homozygous genomes can be obtained by repeated inbreeding, either naturally in species which are self-fertile such as many plants and hermaphroditic animals including the nematode *C. elegans*, or experimentally as in laboratory mice and Drosophila strains. The recent first complete T2T human genome assembly was obtained from a homozygous cell line CHM13hTERT derived from a rare hydatidiform mole that had duplicated a haploid genome[32].

## [H2] Assembling homozygous genomes

For a homozygous genome, the most reliable solution to a near telomere-to-telomere assembly uses both PacBio HiFi reads and ONT ultra-long reads (**Fig. 1a**). We use HiFi reads to construct an initial assembly graph that consists of linear segments (unitigs) which contain no long exact repeats, with possible connections between them depending on the repeat structure. Here "long" is approximately 10kb, with the details dependent on the algorithm used. Highly repetitive regions become represented as complex subgraphs, which we will refer to as tangles. We can then anchor ultra-long reads to the unitigs and thread them through the tangles to resolve most of the tangles. Ultra-long reads can also patch assembly gaps caused by occasional HiFi coverage drops. The human T2T-CHM13 genome was assembled without additional long-range data[20]. Nonetheless, when chromosomes are not separated well or not contiguous in the assembly graph, additional Hi-C data will help to generate chromosome-long scaffolds.

At present, verkko[15,33] and hifiasm[11,12,16] can integrate PacBio HiFi and ONT ultra-long data. They broadly follow the workflow in **Fig. 1a** but use different algorithms at each step. It is possible to achieve good assembly of a homozygous genome with HiFi data alone. Verkko, hifiasm, HiCanu[10] and LJA[13] can all assemble several human chromosomes from telomere to telomere with HiFi reads alone. When scaffolding is necessary, YaHS[34] has replaced SALSA[35] and become the recommended method for high-quality HiFi assemblies with Hi-C data. It is the scaffolder of choice used by both the Vertebrate Genome Project (VGP)[19] and the Darwin Tree of Life project (DToL)[36].

## [H2] Assembling heterozygous diploid genomes

Assembling a heterozygous diploid genome follows a similar strategy (**Fig. 1b**). For genomes with long homozygous regions, including human genomes, the combination of HiFi and ultra-long alone may not phase the whole chromosomes. In this case, trio data that provides accurate phasing across the whole genome is recommended. When it is not possible to obtain parental samples, we may use Hi-C instead. Hi-C only provides relative phasing information between contigs and is not as powerful as trio data especially in tangled subgraphs. Nonetheless, Hi-C remains a key data type for reliably scaffolding chromosomes. VGP and DToL routinely generate Hi-C for most of the species they sequence.

For the time being, ONT ultra-long data is relatively expensive to obtain and requires large amounts for input DNA (typically tens of micrograms). Many sequencing projects do not generate ultra-long data. With HiFi alone, we can produce a primary/alternate assembly pair[37] or a dual assembly pair[12,38] (**Fig. 2**). For deriving a single reference, the primary assembly may be preferred as primary contigs are generally longer. The alternate assembly is fragmented and error prone; it is usually ignored in downstream analysis. Representing both genomes in a diploid sample, the dual assembly pair supports assembly-based variant calling[39] and also the use of both haplotypes in pangenome construction[18]. However, with shorter contigs, scaffolding can be more complex. In either approach there tend to be problems distinguishing paralogous tandem duplications from homologous haplotypic duplications, particularly at the ends of contigs, which can lead to false duplications[40]. For the primary assembly approach, many but not all of these issues can be found and fixed with heuristic methods, as implemented for example in purge_dups[41].

Combining HiFi with long-range data such as trio, Hi-C or Strand-seq, we can produce a pair of haplotype-resolved assemblies[11,12,42,43] (**Fig. 2**). This assembly pair has comparable contiguity to a dual assembly pair. It additionally preserves phasing and can be further scaffolded into phased chromosomes with Hi-C. Even without parental data, it has been shown that it is possible to identify the parental origin of these chromosomes using imprinted methylation makers[44] if they are known and sufficiently frequent to mark each homologous pair of contigs.

For heterozygous genomes, both verkko and hifiasm can integrate PacBio HiFi, ONT ultra-long and long-range data, and can assemble several human chromosomes haplotype-resolved from telomere to telomere. They also work for HiFi data alone and produce dual or primary assembly. HiCanu can also generate a primary assembly with HiFi data and achieve comparable quality.

All the assemblers mentioned so far are optimized for PacBio HiFi data. They may assemble ONT duplex data of similar accuracy but there is as yet insufficient experience to generalise this across a wide range of organisms. They have not been demonstrated to work with other ONT data types, including the latest simplex reads produced by the v14 chemistry. The Shasta-GFAse pipeline[45,46] is a viable choice for contiguous haplotype-resolved assembly from ONT data alone.

## [H1] Core assembly algorithms

Modern long-read assemblers are graph-based. They construct an assembly graph, either an overlap graph[47,48] or a de Bruijn graph[49,50], from input reads. In this graph, a vertex represents a sequence, and an edge indicates a possible connection inferred from reads. An assembly graph ideally retains all information in reads without redundancy. It is however often nonlinear due to repeats and ploidy. We integrate additional data types or rely on graph traversal to resolve remaining ambiguity in the graph and obtain long linear contigs.

Although there are reviews on the theory of graph-based assembly algorithms, their formulations differ. For the completeness of this review, we will describe here the basic theory. In addition, DNA is double-stranded. As a result, unlike classical graphs in graph theory, assembly graphs are bidirected with each edge having two directions. For simplicity in our exposition, we assume DNA sequence only has one strand. Under this assumption, assembly graphs are directed graphs.

## [H2] Read error correction

While PacBio HiFi and ONT duplex reads are accurate, they are not error-free. The remaining errors are mixed with genetic variants and may impede the correct separation of homologous haplotypes or repeat copies, which would lead to fragmented assemblies. All T2T-capable assemblers attempt to correct sequencing errors on reads. HiCanu, verkko and hifiasm align all reads to each other. For each read, they correct a base if it is rarely seen among other overlapping reads aligned to the same position. LJA constructs an initial assembly graph without error correction, aligns each raw read to the graph and takes the graph path of high k-mer coverage as the corrected read sequence. While HiCanu, Verkko and LJA compress homopolymers in reads and correct reads in the homopolymer-compressed space, hifiasm corrects in the original base space. These assemblers can correct the majority of errors[51].

It is quite possible that, to a large extent, the differences between the results of different assemblers on the same data depend as much on variation in error correction as variation in assembly algorithm, but this is hard to establish because the steps are normally integrated. It would be helpful if method developers separated their error correction step from the assembly step.

## [H2] Assembly with overlap graphs

In an overlap graph, each vertex is a read. We add a directed edge from read $A$ to $B$ if a suffix of $A$ can be aligned to a prefix of $B$; in this case, we say $A$ and $B$ have an overlap. **Fig. 3a** shows how to construct an assembly graph from reads. The example results in a single unitig.

Practical overlap graphs are not this clean and need to be further processed. In particular, there will be overlaps between different repeat copies when the repeat length is longer than the overlap length (**Fig. 3b**). In case of human, if we keep overlaps shorter than a few kb, there will be many overlaps between the ~6kb LINE1 repeats. Nonetheless, given reads longer than LINE1, we expect overlaps involving unique regions to be longer than repetitive overlaps. If a read has two overlaps (e.g., read 2 overlaps with 3 and 4 in **Fig. 3b**), shorter overlaps are more likely to be caused by a repeat and may be cut. Meanwhile, uncorrected sequencing errors may lead to extra "tips" (e.g., read 3 in **Fig. 3b**) or "bubbles". These are also removed during graph cleaning.

Most overlap-based assemblers follow the procedure in **Fig. 3a/3b**. HiCanu[10] and hifiasm[11], the two overlap-based assemblers optimized for HiFi reads, are distinct from the rest in that they only allow perfect overlaps. This apparently minor difference is the main source of their power to distinguish repeat copies (**Fig. 3c**) and phase haplotypes (**Fig. 3d**). In this way they achieve more contiguous and more accurate assembly than older assemblers given accurate long reads[52].

It is worth noting that hifiasm is implementing string graphs[48] to be exact. String graph is an alternative formulation of overlap graph. The two types of graphs can be transformed to each other without loss of information. We omit the theoretical details and take them as the same approach.

## [H2] Assembly with de Bruijn graphs

There are two ways to construct a de Bruijn graph (DBG): node-centric or edge-centric[53] (**Fig. 4a**). In a node-centric graph, $DBGv(k)$, each vertex is a $k$-mer in reads, and we have an edge between two *k*-mers if they overlap by *k*-1 bases. In an edge-centric graph, $DBGe(k)$, each edge is a *k*-mer in reads and each vertex is a (*k*-1)-mer. Mathematically speaking, $DBGe(k+1)$ is a subgraph of $DBGv(k)$, while $DBGv(k)$ is the line graph of $DBGe(k)$. In the literature, both definitions are common. We take a node-centric view in this review due to its connection to overlap graphs.

A node-centric DBG is an overlap graph consisting of *k*-mers at vertices with edges corresponding to *k*-1 bp overlaps. It does not have contained reads or transitive edges and is thus simpler. Strategies used for overlap graphs are often applicable to DBG.

A basic DBG discards information longer than the *k*-mer size. This reduces its power for phasing and repeat resolution. It may be tempting to choose a large *k* to retain long-range information, but using a large *k* increases the chance of contig breakpoints in low coverage regions (**Fig. 4b**). There is no single best *k*-mer size for all situations. The multiplex DBG[54,55] provides a good solution to the dilemma of *k*-mer selection. Conceptually, a multiplex DBG can be thought as the merger of multiple DBGs constructed with different *k*-mer sizes (**Fig. 4b**). It adaptively chooses large *k* in repetitive regions and small *k* in low-coverage regions. Nonetheless, using a multiplex DBG does not resolve all the ambiguities in DBGs (**Fig. 4c**). Practical assemblers heuristically use different sets of *k*-mer sizes in different subgraphs[13,15]. They retrieve reads used in a subgraph and replace the subgraph with a new subgraph constructed with longer *k*-mers if the new subgraph is simpler and remains contiguous. This procedure is closer to "read threading"[50] than to the algorithm demonstrated in **Fig. 4b**.

Minimizer-based sparsification[33,56] is another technique employed in modern assemblers. Instead of storing every *k*-mer in reads, we only keep minimizers[57] or closed syncmers[58], a small subset of all *k*-mers, in memory. This strategy greatly reduces memory and speeds up construction. A related but distinct construction called the minimizer-space DBG (mDBG)[14,59] uses *k* consecutive minimizers as "*k*-mers" to construct a DBG. MetaMDBG[59] implements multiplex mDBG which is close to the conceptual definition of multiplex DBG in **Fig. 4b**.

While most short-read assemblers use the DBG approach, no assemblers use DBG for assembling noisy long reads of >5% error rate because long *k*-mers that are required to retain long-range information are mostly wrong. Nonetheless, with accurate long reads, we can correct most sequencing errors away and use *k*-mers over 10 kb in length. DBG once again becomes a viable choice. Verkko[15,33] and LJA[13] are DBG-based assemblers that can achieve broadly comparable assembly quality to HiCanu and hifiasm.

**[H2] Integrating multiple data types**

For a diploid sample, we can phase two heterozygous positions only if there is a read that harbors both sites. The diploid genomes of humans and many other species contain many regions that are longer than HiFi reads and do not contain any heterozygous loci. We would not be able to phase a human genome with HiFi reads alone. At the same time, segmental duplications that happened within the last thousands of years are likely to remain identical over tens of kb and would not be resolved by HiFi reads, either. We need additional data types to get chromosome-long phasing and to assemble recent duplications.

The most powerful auxiliary data type is ONT ultra-long reads, simply owing to their long read length. Verkko aligns ultra-long reads to the initial assembly graph[60] and identifies paths through the graph. It then simplifies the initial graph with the same read threading algorithm for constructing multiple DBG. Hifiasm instead does ultra-long-to-graph alignment with an algorithm similar to minigraph[61]. It encodes an ultra-long read as a sequence of unitigs and applies the overlap-based assembly algorithm in the unitig space. Both assemblers can achieve assembly of higher quality by integrating ultra-long data, but they still cannot phase an entire chromosome.

Trio data, which is obtained by sequencing both parents of a sample, is the most reliable data type for whole-genome phasing[11,25,26]. With a trio, we identify *k*-mers only occurring in one of the parents and use these *k*-mers to mark the parental origin of the unitigs in the assembly graph. We could, for example, connect paternal unitigs and unmarked unitigs to get long paternal contigs. Both verkko and hifiasm can use this sort of information. In practice, however, parental data may be difficult to obtain due to ethical concerns in humans or because the parents are not available for wild-caught animals. In principle data from other close relatives can provide similar information but this type of information is rarely used for assembling new species.

The most common auxiliary data type is Hi-C. It plays two key roles in assembly: scaffolding and phasing. A Hi-C fragment connects two loci if they are spatially close in a nucleus. Due to chromosome packing, loci closer on the same chromosome are more likely to interact in 3D space as well. As a result, if the density of Hi-C reads is high between a pair of contigs, the pair is likely to be close on the same chromosome. Scaffolders use this information to group and order contigs into chromosomes.

Meanwhile, because we are more likely to see Hi-C fragments bridging two loci on the same haploid chromosome than on different homologous chromosomes, Hi-C also provides long-range phasing information. Hifiasm and GFAse adapted reference-based Hi-C phasing algorithms[62,63] for de novo assemblies. Briefly, contigs that harbor the same Hi-C fragments attract each other to the same phasing group; contigs that have high sequence similarity repel each other to opposite phasing groups. Hifiasm and GFAse attempt to find a balance between attractive and repulsive forces to phase contigs. Verkko can optionally take GFAse phasing and generate a chromosome-phased assembly.

## [H1] Evaluating sequence assemblies

For an assembly to be truly telomere-to-telomere it must both cover the whole of each chromosome without gaps, and also not contain large-scale assembly errors. It is critical to rigorously assess the quality of the assembly before concluding it is telomere-to-telomere.

### [H2] Basic metrics

To get a first impression of an assembly, we calculate the assembly size, the sum of all contig lengths, and N50, defined as the length for which contigs no shorter than this number cover half of the assembly. For the autosomes of a diploid sample, we expect the two assemblies in a pair of dual (**Fig. 2c**) or chromosome-phased assemblies (**Fig. 2d**) to have similar sizes. A pair of unbalanced autosomal assemblies may indicate incomplete phasing and may benefit from manual parameter tuning or curation[19]. Of course, in the heterogametic sex (e.g. XY males in mammals) the sex chromosomes will most likely have different sizes. Other ploidy variation within species can also occur, for example due to somatic chromosome loss or diminution[64].

### [H2] Evaluating gene completeness

BUSCO[65] remains a gold standard for evaluating the completeness of an assembly. It works by aligning conserved single-copy proteins to the genome and counting alignments that are missing, broken or duplicated. A less complete assembly would have more proteins unaligned. A caveat is that BUSCO may underestimate the completeness of large genomes. For example, running BUSCO on annotated human genes will lead to a completeness of 99.2%, but running BUSCO on the human genome sequence will result in a completeness of 95.7%. This difference is caused by the difficulty in aligning proteins to the human genome. Compleasm[66] is a recent reimplementation of BUSCO that addresses this problem with more accurate protein-to-genome alignment[67].

The "asmgene" tool from the minimap2 package[68] is an alternative to BUSCO and also solves the low completeness problem when a high-quality reference genome is present. This tool identifies single-copy genes based on cDNA-to-reference alignment. It regards a gene to be complete if it is aligned about equally well to the target assembly. The "asmgene" tool additionally evaluates whether a multi-copy gene in the reference genome is assembled to multiple copies in the target assembly. Noisy read assemblers may have single-copy genes assembled but often miss multi-copy genes[11].

[H2] *K*-mer based evaluation

Given uniform read coverage and a perfect assembly of the reads, we expect the count of a *k*-mer in the assembly to be proportional to its count in reads. A *k*-mer occurring more often in assembly than in reads suggests a false duplication in the assembly, while a *k*-mer having high frequency in the reads but absent from the assembly suggests missing sequences. KAT[69] is a powerful tool that makes use of these simple observations to evaluate an assembly.

It is now a common practice to use *k*-mers to estimate the base accuracy of contig sequences, often measured in the Phred scale[70] as QV (Quality Value). This method works by calculating the fraction of contig *k*-mers that are absent from reads. A higher fraction suggests a lower QV. Currently there are two implementations, Merqury[71] and yak[11]. Yak is different in that it attempts to correct overestimated QV at high read depth. It is worth noting that because QV estimation depends on the quality and the depth of input reads, QV estimates based on different input reads are not strictly comparable. We cannot easily conclude the QV of one species to be higher than the QV of another species. When there is trio data, both Merqury and yak can also use parent-specific *k*-mers to evaluate the phasing accuracy of an assembly.

[H2] Alignment-based evaluation

Ideally, when we align sequence reads to their assembly, we expect even coverage at every contig position. Excessively low or high coverage over a long region would indicate a potential assembly error. We also expect contigs to be well supported by reads at base level. When we call variants from the read-to-assembly alignment, isolated confident small variant calls would indicate contig consensus errors, while clustered heterozygous variant calls could result from collapsed segmental duplications. Such signals played a crucial role in evaluating the homozygous CHM13 genome. For a diploid genome, we may merge the two haplotype assemblies and map reads to the merge. We should observe similar signals. Flagger, Asset[72] and Inspector[73] are user-facing evaluation tools based on read-to-assembly alignment.

For a sample with a near perfect curated assembly, such as CHM13, we can take the existing assembly as the ground truth to evaluate automated assemblies produced with fewer data types or at lower read coverage. QUAST[74] is a popular tool for this purpose. Such methods based on assembly-to-assembly alignment are invaluable for assembler developers to tune assembly algorithms but are not applicable to new species or when the "truth" assembly and the evaluated assembly were derived from different strains or different samples. In the latter case, QUAST would often report structural variants as misassemblies. An assembly more complete in complex regions would appear to have higher error rate. For example, if we take the human reference genome GRCh38 as ground truth to evaluate the telomere-to-telomere CHM13 assembly, QUAST would report over 20,000 misassemblies[11].

## [H1] Challenges in de novo sequence assembly

Despite the progress outlined above, de novo sequence assembly is not a solved problem. While all mainstream assemblers are built upon basic assembly algorithms established by 1995, they heavily rely on hand tuned heuristics that do not have a solid theoretical foundation. Limited by the characteristics of practical data, they cannot resolve the most complex regions in genomes. They also perform poorly with polyploid genomes or more complex cases such as cancer genomes with heterogeneous large scale structural variation.

### [H2] Theoretical challenges

Of the two assembly paradigms, overlap graph and de Bruijn graph, each has its own caveats. When constructing an overlap graph, we discard a read contained in longer reads. This apparently straightforward step may lead to assembly gaps when reads are variable in length (**Fig. 3e**). Such assembly gaps are infrequent but as modern assemblies are highly contiguous, additional assembly gaps caused by contained reads are noticeable. To alleviate this problem, hifiasm tries to rescue a contained read if having the read would patch an assembly gap. This heuristic works in simple cases but is not always reliable. Containment removal is the Achilles' heel of overlap-based assembly algorithms[75,76] and remains an open and critical problem.

Leading DBG assemblers, including SPAdes, Verkko and LJA, all use multiplex DBG. While constructing multiplex DBG from a fixed set of *k*-mers across all input reads is well studied, modern DBG assemblers are not using these algorithms because one *k*-mer set may not work optimally across all subgraphs (**Fig. 4c**). They resort to heuristics and walk on the fine line between graph contiguity and complexity, not governed by solid theory.

Multiplex DBG as it is implemented in modern assemblers is distinct from the basic DBG described in textbooks. It will be interesting to see if we could go a step further and come up with a new assembly paradigm to combine different length scales more smoothly.

### [H2] Practical challenges

Assembly using Hi-C as long-range data is more difficult than assembly using trios. Without trios, current assemblers may have troubles with phasing acrocentric chromosomes or micro-chromosomes. For a sample having different sex chromosomes (e.g., a male in a XY system or a female in a ZW system), they often do not cleanly separate the sex chromosomes, either. Nonetheless, by inspecting Hi-C alignment, human curators can often identify these issues and manually fix them. This suggests there is further room for improvement in using Hi-C data, perhaps using machine learning approaches.

The current telomere-to-telomere strategy emphasizes the use of long reads with accuracy well above 99%. In fact, they rely on pre-assembly error correction to operate so that the majority of reads are perfect. With the latest v14 ONT chemistry, ONT simplex reads can reach an accuracy 98–99%. We suspect it may be possible to devise error correction strategies for the new ONT simplex reads that allow use of current or adapted long read assembly algorithms that require exact matches. If this works, there is a prospect of accurate telomere-to-telomere assembly from a single data type, which would greatly streamline high quality genome assembly.

Experimental challenges often have a bigger impact than computational challenges in practice. Standard protocols for long read generation require large amounts of DNA (over a microgram) which can be hard or impossible to obtain from small organisms or clinical samples. There are unamplified low input library protocols for PacBio that work down to 0.1 microgram[77], but below this whole genome amplification is necessary, which introduces coverage bias and dropouts. There are also coverage biases with long reads. For example, PacBio HiFi reads struggle to sequence through long GA-rich repeats. ONT reads have trouble with telomere sequences[78]. ONT duplex reads may also have sequence-dependent coverage drops. These coverage biases were shown to lead to over 25% of the assembly gaps in a recent study[79].

**[H2] Beyond diploid samples**

While excellent progress has been made recently on telomere-to-telomere assembly of diploid genomes, we do not have a satisfactory solution for polyploid genomes. There have been two recent haplotype phased assemblies of the tetraploid potato, using single-cell sequencing data[80] or genetic maps[81] for phasing, but these are not methods that can be deployed at scale. We would prefer to derive polyploid assembly using common data types; in principle the necessary information should be present in HiFi long reads, ultra-long reads and Hi-C data.

A cancer genome is polyploid to some extent, with ploidy varying between or within chromosomes. They are even harder to assemble. Beyond cancer, we would like to assemble metagenomic samples which contain a large variety of species, typically microbial, at very different relative abundances. A metagenome can also be considered as a polyploid genome with even higher ploidy variation. Nonetheless, when we assemble a metagenome sample, the bar is lower in comparison to polyploid genome assembly: for example, it is not normally expected to phase highly similar genomes. There are dedicated metagenome assemblers, such as MetaFlye[82], hifiasm-meta[83] and metaMDBG[59], that can reconstruct up to a few hundred closed bacterial genomes from a deeply sequenced metagenome sample, although they are still missing species detectable based on 16S or *k*-mer profiling[84]. There is still a long way to go to achieve complete metagenome assembly.

**[H1] Conclusions/Outlook**

Thanks to the availability of PacBio HiFi reads and ONT ultra-long reads, the quality of de novo assembly has improved dramatically in the past two years. Now a fully automated assembler can phase and assemble some chromosomes from telomere to telomere for diploid mammals and other species with large genomes. This was unthinkable in mid 2020.

Can we automatically assemble all chromosomes from telomere to telomere with current data? We think the answer is generally "no". We believe that most of the advances in the past few years have been made because of improvements in data quality, and current assemblers pull most of the information from the available input data. Algorithm improvement alone may not reliably resolve all assembly gaps. We look forward to continuing new advances in sequencing technologies to truly complete a genome without human intervention.

It is important to note that a complete assembly only sets a start for downstream biological discoveries. While genome assembly has progressed rapidly, genome alignment and annotation tools have lagged far behind. We hope to see continued development of these tools in the future to realize the full power of (near) complete assembly.

## [H1] References

1. C. elegans Sequencing Consortium. Genome sequence of the nematode C. elegans: a platform for investigating biology. *Science* **282**, 2012–2018 (1998).
2. Schneider, V. A. *et al.* Evaluation of GRCh38 and de novo haploid genome assemblies demonstrates the enduring quality of the reference assembly. *Genome Res.* **27**, 849–864 (2017).
3. Lander, E. S. *et al.* Initial sequencing and analysis of the human genome. *Nature* **409**, 860–921 (2001).
4. Myers, E. W. *et al.* A whole-genome assembly of Drosophila. *Science* **287**, 2196–2204 (2000).
5. Venter, J. C. *et al.* The sequence of the human genome. *Science* **291**, 1304–1351 (2001).
6. Bentley, D. R. *et al.* Accurate whole human genome sequencing using reversible terminator chemistry. *Nature* **456**, 53–59 (2008).
7. Chin, C.-S. *et al.* Nonhybrid, finished microbial genome assemblies from long-read SMRT sequencing data. *Nat. Methods* **10**, 563–569 (2013).
8. Chaisson, M. J. P. *et al.* Resolving the complexity of the human genome using single-molecule sequencing. *Nature* **517**, 608–611 (2015).
9. Wenger, A. M. *et al.* Accurate circular consensus long-read sequencing improves variant detection and assembly of a human genome. *Nat. Biotechnol.* **37**, 1155–1162 (2019).
10. Nurk, S. *et al.* HiCanu: accurate assembly of segmental duplications, satellites, and allelic variants from high-fidelity long reads. *Genome Res.* **30**, 1291–1305 (2020).
11. Cheng, H., Concepcion, G. T., Feng, X., Zhang, H. & Li, H. Haplotype-resolved de novo assembly using phased assembly graphs with hifiasm. *Nat. Methods* **18**, 170–175 (2021).
12. Cheng, H. *et al.* Haplotype-resolved assembly of diploid genomes without parental data. *Nat. Biotechnol.* **40**, 1332–1335 (2022).
13. Bankevich, A., Bzikadze, A. V., Kolmogorov, M., Antipov, D. & Pevzner, P. A. Multiplex de Bruijn graphs enable genome assembly from long, high-fidelity reads. *Nat. Biotechnol.* **40**, 1075–1081 (2022).
14. Ekim, B., Berger, B. & Chikhi, R. Minimizer-space de Bruijn graphs: Whole-genome assembly of long reads in minutes on a personal computer. *Cell Syst.* **12**, 958-968.e6 (2021).
15. Rautiainen, M. *et al.* Telomere-to-telomere assembly of diploid chromosomes with Verkko. *Nat. Biotechnol.* (2023) doi:10.1038/s41587-023-01662-6.
16. Cheng, H., Asri, M., Lucas, J., Koren, S. & Li, H. Scalable telomere-to-telomere assembly for diploid and polyploid genomes with double graph. (2023) doi:10.48550/ARXIV.2306.03399.
17. Liao, W.-W. *et al.* A draft human pangenome reference. *Nature* **617**, 312–324 (2023).
18. Gao, Y. *et al.* A pangenome reference of 36 Chinese populations. *Nature* **619**, 112–121 (2023).

19. Rhie, A. *et al.* Towards complete and error-free genome assemblies of all vertebrate species. *Nature* **592**, 737–746 (2021).
20. Nurk, S. *et al.* The complete sequence of a human genome. *Science* **376**, 44–53 (2022).
21. Altemose, N. *et al.* Complete genomic and epigenetic maps of human centromeres. *Science* **376**, eabl4178 (2022).
22. Burton, J. N. *et al.* Chromosome-scale scaffolding of de novo genome assemblies based on chromatin interactions. *Nat. Biotechnol.* **31**, 1119–1125 (2013).
23. Deshpande, A. S. *et al.* Identifying synergistic high-order 3D chromatin conformations from genome-scale nanopore concatemer sequencing. *Nat. Biotechnol.* **40**, 1488–1499 (2022).
24. Falconer, E. *et al.* DNA template strand sequencing of single-cells maps genomic rearrangements at high resolution. *Nat. Methods* **9**, 1107–1112 (2012).
25. Malinsky, M., Simpson, J. T. & Durbin, R. *trio-sga : facilitating* de novo *assembly of highly heterozygous genomes with parent-child trios*. http://biorxiv.org/lookup/doi/10.1101/051516 (2016) doi:10.1101/051516.
26. Koren, S. *et al.* De novo assembly of haplotype-resolved genomes with trio binning. *Nat. Biotechnol.* (2018) doi:10.1038/nbt.4277.
27. Wang, O. *et al.* Efficient and unique cobarcoding of second-generation sequencing reads from long DNA molecules enabling cost-effective and accurate sequencing, haplotyping, and de novo assembly. *Genome Res.* **29**, 798–808 (2019).
28. Chen, Z. *et al.* Ultralow-input single-tube linked-read library method enables short-read second-generation sequencing systems to routinely generate highly accurate and economical long-range sequencing information. *Genome Res.* **30**, 898–909 (2020).
29. Meier, J. I. *et al.* Haplotype tagging reveals parallel formation of hybrid races in two butterfly species. *Proc. Natl. Acad. Sci. U. S. A.* **118**, e2015005118 (2021).
30. Miga, K. H. *et al.* Telomere-to-telomere assembly of a complete human X chromosome. *Nature* **585**, 79–84 (2020).
31. Bzikadze, A. V. & Pevzner, P. A. Automated assembly of centromeres from ultra-long error-prone reads. *Nat. Biotechnol.* **38**, 1309–1316 (2020).
32. Taillon-Miller, P. *et al.* The homozygous complete hydatidiform mole: a unique resource for genome studies. *Genomics* **46**, 307–310 (1997).
33. Rautiainen, M. & Marschall, T. MBG: Minimizer-based sparse de Bruijn Graph construction. *Bioinformatics* **37**, 2476–2478 (2021).
34. Zhou, C., McCarthy, S. A. & Durbin, R. YaHS: yet another Hi-C scaffolding tool. *Bioinformatics* **39**, btac808 (2023).
35. Ghurye, J. *et al.* Integrating Hi-C links with assembly graphs for chromosome-scale assembly. *PLoS Comput. Biol.* **15**, e1007273 (2019).
36. Darwin Tree of Life Project Consortium. Sequence locally, think globally: The Darwin Tree of Life Project. *Proc. Natl. Acad. Sci. U. S. A.* **119**, e2115642118 (2022).
37. Chin, C.-S. *et al.* Phased diploid genome assembly with single-molecule real-time sequencing. *Nat. Methods* **13**, 1050–1054 (2016).
38. Chin, C.-S. & Khalak, A. *Human Genome Assembly in 100 Minutes*. http://biorxiv.org/lookup/doi/10.1101/705616 (2019) doi:10.1101/705616.
39. Li, H. *et al.* A synthetic-diploid benchmark for accurate variant-calling evaluation. *Nat. Methods* **15**, 595–597 (2018).
40. Ko, B. J. *et al.* Widespread false gene gains caused by duplication errors in genome assemblies. *Genome Biol.* **23**, 205 (2022).
41. Guan, D. *et al.* Identifying and removing haplotypic duplication in primary genome assemblies. *Bioinformatics* **36**, 2896–2898 (2020).
42. Garg, S. *et al.* Chromosome-scale, haplotype-resolved assembly of human genomes. *Nat. Biotechnol.* **39**, 309–312 (2021).

43. Porubsky, D. *et al.* Fully phased human genome assembly without parental data using single-cell strand sequencing and long reads. *Nat. Biotechnol.* **39**, 302–308 (2021).
44. Akbari, V. *et al.* Parent-of-origin detection and chromosome-scale haplotyping using long-read DNA methylation sequencing and Strand-seq. *Cell Genomics* **3**, 100233 (2023).
45. Shafin, K. *et al.* Nanopore sequencing and the Shasta toolkit enable efficient de novo assembly of eleven human genomes. *Nat. Biotechnol.* **38**, 1044–1053 (2020).
46. Lorig-Roach, R. *et al. Phased nanopore assembly with Shasta and modular graph phasing with GFAse*. http://biorxiv.org/lookup/doi/10.1101/2023.02.21.529152 (2023) doi:10.1101/2023.02.21.529152.
47. Myers, E. W. Toward simplifying and accurately formulating fragment assembly. *J Comput Biol* **2**, 275–290 (1995).
48. Myers, E. W. The fragment assembly string graph. *Bioinformatics* **21 Suppl 2**, ii79-85 (2005).
49. Idury, R. M. & Waterman, M. S. A new algorithm for DNA sequence assembly. *J Comput Biol* **2**, 291–306 (1995).
50. Pevzner, P. A., Tang, H. & Waterman, M. S. An Eulerian path approach to DNA fragment assembly. *Proc. Natl. Acad. Sci. U. S. A.* **98**, 9748–9753 (2001).
51. Guo, Y., Feng, X. & Li, H. *Evaluation of haplotype-aware long-read error correction with hifieval*. http://biorxiv.org/lookup/doi/10.1101/2023.06.05.543788 (2023) doi:10.1101/2023.06.05.543788.
52. Jarvis, E. D. *et al.* Semi-automated assembly of high-quality diploid human reference genomes. *Nature* **611**, 519–531 (2022).
53. Chikhi, R., Limasset, A. & Medvedev, P. Compacting de Bruijn graphs from sequencing data quickly and in low memory. *Bioinformatics* **32**, i201–i208 (2016).
54. Peng, Y., Leung, H. C. M., Yiu, S. M. & Chin, F. Y. L. IDBA-UD: a de novo assembler for single-cell and metagenomic sequencing data with highly uneven depth. *Bioinformatics* **28**, 1420–1428 (2012).
55. Bankevich, A. *et al.* SPAdes: a new genome assembly algorithm and its applications to single-cell sequencing. *J Comput Biol* **19**, 455–477 (2012).
56. Ye, C., Ma, Z. S., Cannon, C. H., Pop, M. & Yu, D. W. Exploiting sparseness in de novo genome assembly. *BMC Bioinformatics* **13 Suppl 6**, S1 (2012).
57. Roberts, M., Hayes, W., Hunt, B. R., Mount, S. M. & Yorke, J. A. Reducing storage requirements for biological sequence comparison. *Bioinformatics* **20**, 3363–3369 (2004).
58. Edgar, R. Syncmers are more sensitive than minimizers for selecting conserved k-mers in biological sequences. *PeerJ* **9**, e10805 (2021).
59. Benoit, G. *et al. Efficient High-Quality Metagenome Assembly from Long Accurate Reads using Minimizer-space de Bruijn Graphs*. http://biorxiv.org/lookup/doi/10.1101/2023.07.07.548136 (2023) doi:10.1101/2023.07.07.548136.
60. Rautiainen, M. & Marschall, T. GraphAligner: rapid and versatile sequence-to-graph alignment. *Genome Biol.* **21**, 253 (2020).
61. Li, H., Feng, X. & Chu, C. The design and construction of reference pangenome graphs with minigraph. *Genome Biol.* **21**, 265 (2020).
62. Edge, P., Bafna, V. & Bansal, V. HapCUT2: robust and accurate haplotype assembly for diverse sequencing technologies. *Genome Res.* **27**, 801–812 (2017).
63. Tourdot, R. W., Brunette, G. J., Pinto, R. A. & Zhang, C.-Z. Determination of complete chromosomal haplotypes by bulk DNA sequencing. *Genome Biol.* **22**, 139 (2021).
64. Du, K. *et al.* The sterlet sturgeon genome sequence and the mechanisms of segmental rediploidization. *Nat. Ecol. Evol.* **4**, 841–852 (2020).

65. Manni, M., Berkeley, M. R., Seppey, M., Simão, F. A. & Zdobnov, E. M. BUSCO Update: Novel and Streamlined Workflows along with Broader and Deeper Phylogenetic Coverage for Scoring of Eukaryotic, Prokaryotic, and Viral Genomes. *Mol. Biol. Evol.* **38**, 4647–4654 (2021).
66. Huang, N. & Li, H. *miniBUSCO: a faster and more accurate reimplementation of BUSCO*. http://biorxiv.org/lookup/doi/10.1101/2023.06.03.543588 (2023) doi:10.1101/2023.06.03.543588.
67. Li, H. Protein-to-genome alignment with miniprot. *Bioinformatics* **39**, btad014 (2023).
68. Li, H. Minimap2: pairwise alignment for nucleotide sequences. *Bioinformatics* **34**, 3094–3100 (2018).
69. Mapleson, D., Garcia Accinelli, G., Kettleborough, G., Wright, J. & Clavijo, B. J. KAT: a K-mer analysis toolkit to quality control NGS datasets and genome assemblies. *Bioinformatics* **33**, 574–576 (2017).
70. Ewing, B. & Green, P. Base-calling of automated sequencer traces using phred. II. Error probabilities. *Genome Res.* **8**, 186–194 (1998).
71. Rhie, A., Walenz, B. P., Koren, S. & Phillippy, A. M. Merqury: reference-free quality, completeness, and phasing assessment for genome assemblies. *Genome Biol.* **21**, 245 (2020).
72. Guan, D. *et al. Genome sequence assembly evaluation using long-range sequencing data*. http://biorxiv.org/lookup/doi/10.1101/2022.05.10.491304 (2022) doi:10.1101/2022.05.10.491304.
73. Chen, Y., Zhang, Y., Wang, A. Y., Gao, M. & Chong, Z. Accurate long-read de novo assembly evaluation with Inspector. *Genome Biol.* **22**, 312 (2021).
74. Mikheenko, A., Prjibelski, A., Saveliev, V., Antipov, D. & Gurevich, A. Versatile genome assembly evaluation with QUAST-LG. *Bioinformatics* **34**, i142–i150 (2018).
75. Hui, J., Shomorony, I., Ramchandran, K. & Courtade, T. A. Overlap-based genome assembly from variable-length reads. in *2016 IEEE International Symposium on Information Theory (ISIT)* 1018–1022 (IEEE, 2016). doi:10.1109/ISIT.2016.7541453.
76. Jain, C. Coverage-preserving sparsification of overlap graphs for long-read assembly. *Bioinformatics* **39**, btad124 (2023).
77. Lawniczak, M. K. N. *et al.* Standards recommendations for the Earth BioGenome Project. *Proc. Natl. Acad. Sci. U. S. A.* **119**, e2115639118 (2022).
78. Tan, K.-T., Slevin, M. K., Meyerson, M. & Li, H. Identifying and correcting repeat-calling errors in nanopore sequencing of telomeres. *Genome Biol.* **23**, 180 (2022).
79. Porubsky, D. *et al.* Gaps and complex structurally variant loci in phased genome assemblies. *Genome Res.* **33**, 496–510 (2023).
80. Sun, H. *et al.* Chromosome-scale and haplotype-resolved genome assembly of a tetraploid potato cultivar. *Nat. Genet.* **54**, 342–348 (2022).
81. Bao, Z. *et al.* Genome architecture and tetrasomic inheritance of autotetraploid potato. *Mol. Plant* **15**, 1211–1226 (2022).
82. Kolmogorov, M. *et al.* metaFlye: scalable long-read metagenome assembly using repeat graphs. *Nat. Methods* **17**, 1103–1110 (2020).
83. Feng, X., Cheng, H., Portik, D. & Li, H. Metagenome assembly of high-fidelity long reads with hifiasm-meta. *Nat. Methods* **19**, 671–674 (2022).
84. Feng, X. & Li, H. Towards complete representation of bacterial contents in metagenomic samples. (2022) doi:10.48550/ARXIV.2210.00098.
85. Naish, M. *et al.* The genetic and epigenetic landscape of the Arabidopsis centromeres. *Science* **374**, eabi7489 (2021).

**[H1] Highlighted references**

XX. Cheng, H., Concepcion, G. T., Feng, X., Zhang, H. & Li, H. Haplotype-resolved de novo assembly using phased assembly graphs with hifiasm. *Nat. Methods* **18**, 170–175 (2021).

**This is the primary paper describing hifiasm, a widely used assembler that produces high-quality assembly by integrating multiple data types.**

XX. Bankevich, A., Bzikadze, A. V., Kolmogorov, M., Antipov, D. & Pevzner, P. A. Multiplex de Bruijn graphs enable genome assembly from long, high-fidelity reads. *Nat. Biotechnol.* **40**, 1075–1081 (2022)

**This paper describes the application of multiplex de Bruijn graph to accurate long read assembly.**

XX. Rautiainen, M. et al. Telomere-to-telomere assembly of diploid chromosomes with Verkko. *Nat. Biotechnol.* (2023) doi:10.1038/s41587-023-01662-6.

**This paper describes verkko that integrates PacBio HiFi and ONT ultra-long data for automated high-quality assembly.**

XX. Rhie, A. et al. Towards complete and error-free genome assemblies of all vertebrate species. *Nature* **592**, 737–746 (2021)

**Presentation of 16 chromosomal assemblies of diverse vertebrate species, highlighting the improvements in assembly quality derived from long read assembly.**

XX. Nurk, S. et al. The complete sequence of a human genome. *Science* **376**, 44–53 (2022).

**Seminal paper giving the first telomere-to-telomere human genome.**

XX. Zhou, C., McCarthy, S. A. & Durbin, R. YaHS: yet another Hi-C scaffolding tool. *Bioinformatics* **39**, btac808 (2023).

**This paper describes current state of the art Hi-C scaffolding method.**

## [H1] Acknowledgements

We are grateful to Haoyu Cheng and Chenxi Zhou for their helpful comments on the manuscript. This study was supported by grants from the US National Institutes of Health (R01HG010040 to H.L.) and from Wellcome Trust (226458 to R.D.).

## [H1] Author contributions

Both authors contributed to all aspects of the review.

## [H1] Competing interests statement

The authors declare no competing interests.

# [H1] Glossary terms

- **Genome**. The set of distinct chromosomal sequences from an organism.
- **Read**. The nucleotide sequence of a fragment of DNA inferred by a sequencing instrument.
- **Assembly**. A set of non-redundant sequences supposedly representing a genome or regions in a genome. Assembly also denotes the process of reconstructing a genome from sequence reads.
- **Contig**. A contiguous sequence in an assembly. A contig does not contain long stretches of unknown sequence.
- **Assembly gap**. A region in the genome that is not assembled into contigs.
- **Scaffold**. A sequence consisting of multiple contigs connected in a defined order and orientation with gaps.
- **Telomere-to-telomere contig or scaffold**. A sequence representing an entire eukaryotic chromosome.
- **Unitig**. A sequence of a non-branching path in an assembly graph.
- **Ploidy**. The number of homologous copies of chromosomes in the cells of an organism. Ploidy can vary between chromosomes, parts of a chromosome, and cells in an organism.
- **Haplotig**. A contig or scaffold that comes from a single haploid chromosome. A contig that is not a haplotig may contain subsequences from two or more homologous chromosomes.
- **Haplotype-resolved assembly**. An assembly composed of haplotigs only.
- **Diploid assembly**. Two sets of non-redundant sequences with each set representing one haploid genome in a diploid sample.

# [H1] Display Items

Table 1 | **Common data types for high-quality assembly**

| Data type | Technologies | Description | Roles |
|---|---|---|---|
| Accurate long reads | PacBio HiFi, ONT duplex | >10 kb in length; error rate <0.5% | Initial assembly graph construction; phasing where variants are <10kb apart |
| Ultra-long reads | ONT ultra-long | >100 kb in length; error rate <10% | Resolving tangles; longer range phasing |
| Trio data | Short-read | Standard WGS of parents | Whole-genome phasing |
| Long-range data | Hi-C, Pore-C, Strand-seq | Information over 1 kb – >10 Mb | chromosomal phasing; chromosome-scale scaffolding |

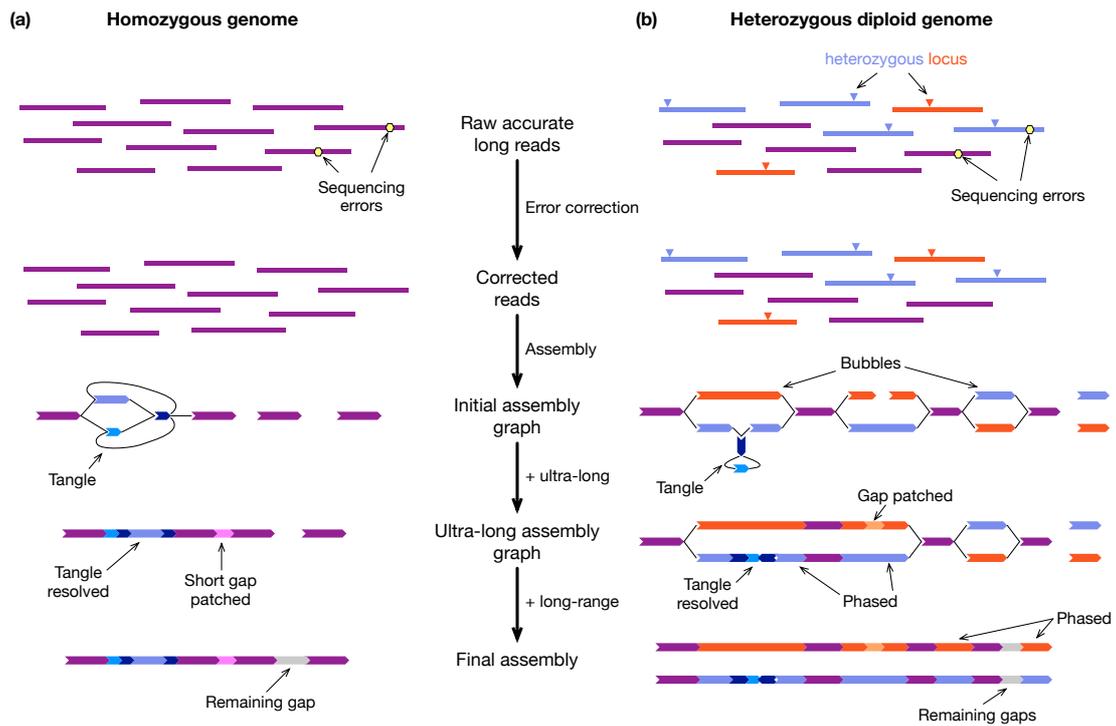

Figure 1| **Strategy for near telomere-to-telomere assembly. a**, Assembling a haploid or homozygous genome. After sequencing errors on accurate long reads are corrected, error-free reads are assembled into an initial assembly graph, where a thick arrow denotes a sequence, and a thin line connects sequences. Ultra-long reads are then threaded through the assembly graph to resolve tangled subgraphs and patch small assembly gaps. Long-range data such as Hi-C helps to scaffold across remaining gaps. **b**, Assembling a heterozygous diploid genome. Heterozygous differences between haplotypes are preserved during error correction. The assembly graphs often consist of a chain of "bubbles", representing polymorphisms between haplotypes. Ultra-long reads and long-range data can be used to phase haplotypes as well as resolve tangles.

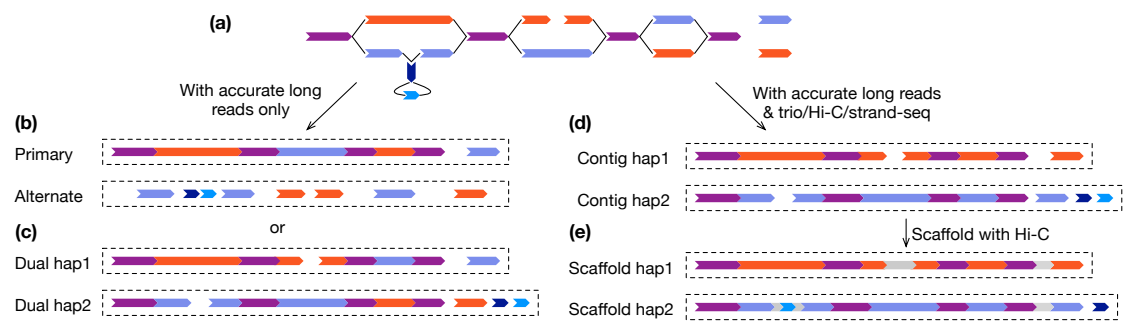

Figure 2| **Types of phased assembly of diploid samples. a**, The assembly graph from **Fig. 1b** can be further processed into different types of assemblies. **b**, Primary/alternate assembly pair. The primary assembly represents a complete haploid genome with occasional phase switches. The alternate assembly is fragmented. **c**, A pair of dual assemblies. Each dual assembly is similar to a primary assembly. **d**, A pair of chromosome-phased assemblies. Contigs from the same haploid chromosome are partitioned to the same assembly. **e**, A pair of chromosome-phased assemblies with scaffolding. Contigs are joined into chromosomes across assembly gaps.

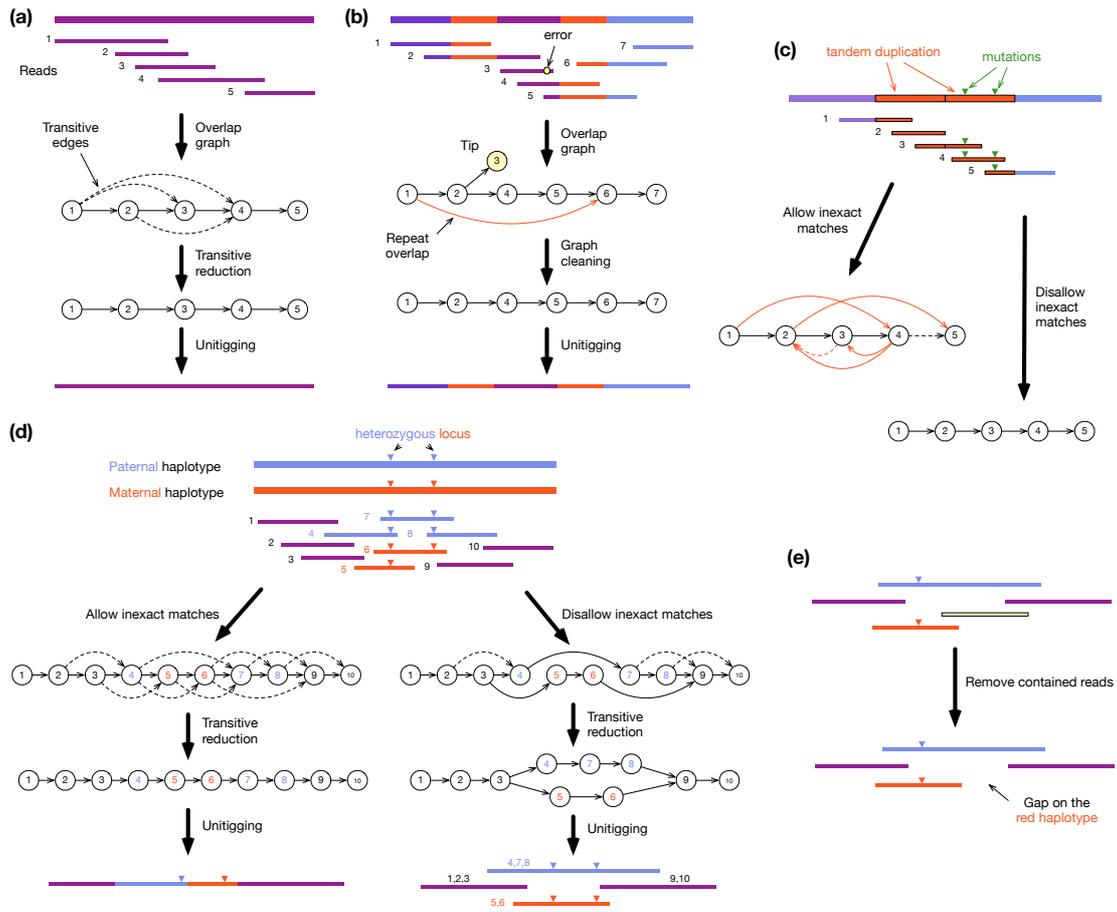

Figure 3| **Assembly with overlap graphs. a**, Simple overlap graph assembly. Find overlaps between all reads, identify transitive overlaps (dashed arrows) that can be inferred from other overlaps, remove transitive overlaps, and merge vertices with one incoming edge and one outgoing edge to get the final unitigs. **b**, Graph cleaning. An uncorrected sequencing error (yellow hexagon) may lead to a tip (read 3) that should be trimmed off. Repeats (red regions) may result in overlaps between repeat copies that can be cut with graph cleaning. **c**, Assembling a tandem duplication longer than reads. Disallowing inexact overlaps (red arrows) resolves the region into a simple graph. **d**, Assembling a diploid sample. Allowing inexact overlaps leads to the loss of heterozygous differences and collapses the two haplotypes. Using only exact overlaps eliminates alignments between haplotypes and thus preserves the heterozygous alleles and their local phasing. **e,** Removing contained reads (yellow lines) leads to assembly gaps on the red haplotypes.

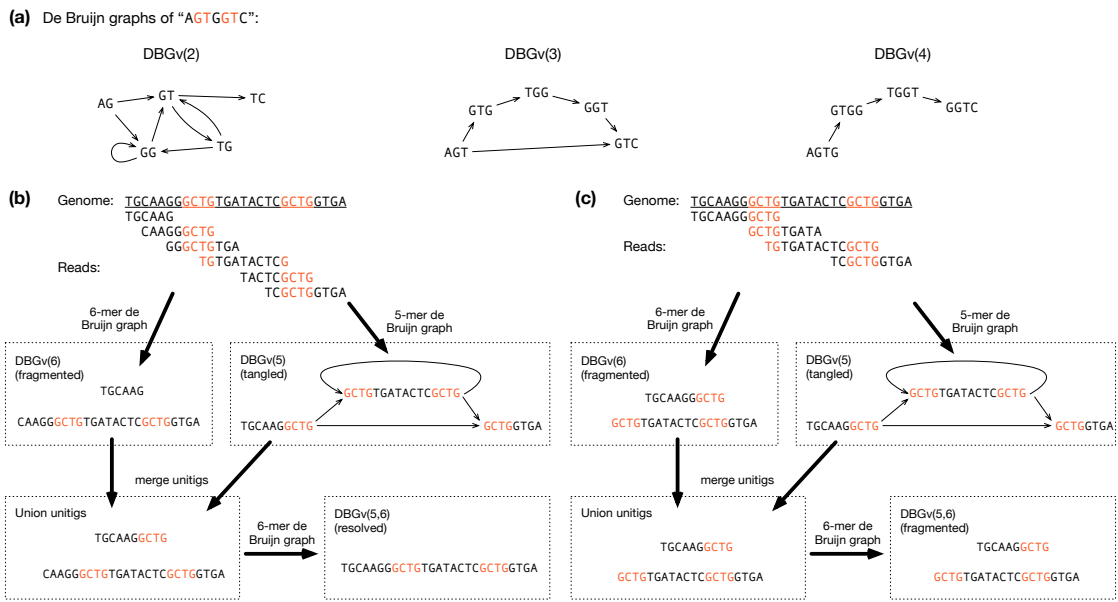

Figure 4 | **De Bruijn graphs. a**, Node(vertex)-centric de Bruijn graphs of a string of different k-mer lengths. **b**, Multiplex DBG improves assembly. The compacted de Bruijn graph using 6-mers as nodes, $DBGv(6)$, is fragmented into two unitgs. $DBGv(5)$ has one connected component but the graph has a cycle. A multiplex de Bruijn graph, $DBGv(5,6)$, is conceptually constructed from the combined set of unitigs in $DBGv(5)$ and $DBGv(6)$, using 6-mers as nodes. **c**, However, multiplex DBG does not resolve all cases. In this case, the multiplex DBG is still fragmented, while an overlap-based method (requiring ≥4bp overlaps) assembles to a single contig (as in **b**).

**Box 1 | Repetitive sequences and de novo assembly**

A repetitive sequence, or repeat, is a sequence that occurs multiple times in the genome. The copies of a repeat do not have to be identical; they may slightly differ from one another. Repetitive sequences are abundant in eukaryotic genomes and are the leading factor that complicates de novo assembly. A repeat can be computationally resolved by a very long read that bridges between non-repetitive sequences on both sides of the repeated region, or by long reads that are accurate enough to distinguish inexact repeat copies (bridging between the differences between the repeats).

Repetitive sequences can be approximately classified into three categories: interspersed repeats, tandem repeats, and segmental duplications. Interspersed repeats are mostly transposable elements scattered in the genome. They are almost all shorter than modern long reads and so no longer pose a major challenge to assembly. Most tandem repeats on chromosome arms are shorter than long reads and hence are easy to assemble as well. However, satellite repeats, a type of extra-long tandem repeat typically enriched in centromeres, are particularly hard to assemble because an entire satellite array cannot be spanned by long reads. Segmental duplications refer to very long DNA segments duplicated in the genome, frequently longer than long or even ultralong reads. Many of them are clustered and can be tandem. While ancient fixed segmental duplications are easy to resolve because they have accumulated differences by mutation since their common ancestor, long polymorphic duplications are challenging. Ribosome DNA (rDNA) may be organized as long tandem arrays consisting of highly similar copies. Long rDNA arrays are among the most difficult regions to assemble[20,85].